\documentclass[twoside]{article}
\usepackage{fleqn,espcrc2,epsfig,graphicx}

\usepackage{graphicx}
\usepackage[figuresright]{rotating}

\newcommand{\AmS}{{\protect\the\textfont2
  A\kern-.1667em\lower.5ex\hbox{M}\kern-.125emS}}

\title{Density response of the t-J model and renormalization
       of breathing and half-breathing phonon modes: A slave-fermion calculation.}

\author{P. Horsch$^a$
        \thanks{Materials and Mechanisms of Superconductivity High Temperature 
         Superconductors VI, Houston, USA, Febr. 20-25,2000; 
         publ. in Physica C 341-348, 117 (2000)}
, G. Khaliullin
        \address{Max-Planck-Institut f\"ur Festk\"orperforschung, 
        Heisenbergstr.1, D-70569 Stuttgart, Germany}%
        \thanks{Permanent address: Kazan Physicotechnical Institute,
                 420029 Kazan, Russia.}
        and 
        V. Oudovenko$^a$\address{Serin Physics Laboratory, Rutgers University, 
        Piscataway, New Jersey 08855-0849, USA.    }
}
       
\begin{document}

\begin{abstract}
The density fluctuation spectrum $N({\bf k},\omega)$ is calculated for the $t$-$J$
model in the low-doping regime using a slave-fermion method for the constrained 
fermions.
The obtained results for  $N({\bf k},\omega)$ are in good agreement with 
diagonalization results. The density response is characterized by incoherent, 
momentum dependent spectral functions reaching up to energies $\sim 8t$ and
a low-energy structure at energy $\sim J$ due to transitions in the 
quasiparticle band.
$N({\bf k},\omega)$ is shown to lead to a strong renormalization
of planar bond-streching and breathing phonon modes with a 
large phonon linewidth at intermediate momenta caused by the low-energy response.
Our results are consistent with recent neutron scattering data, showing 
the peculiar behavior of these modes.
\vspace{1pc}
\end{abstract}

\maketitle

\section{INTRODUCTION}

The density response of doped Mott-Hubbard insulators like the high-T$_c$ 
superconductors is expected to differ strongly from that of weakly correlated 
metals.   
Due to the Mott-Hubbard gap the low-energy density response 
in the doped systems is proportional to the number of holes
and not proportional to the electron concentration.
Experimentally the peculiarities of the density response in high-temperature
superconductors show up (i) in the anomalous mid-infrared absorption in the
frequency dependent conductivity, (ii) in the renormalization and 
strong increase of linewidth of certain 
phonon modes upon doping, and 
(iii) the charge instability leading to the stripe phase. 
While static stripes in cuprates are found only in the LTT-phase 
it is expected that dynamic stripe like correlations exist also in the other
cuprate materials.

Exact diagonalization studies\cite{toh95} 
of the dynamical density response $N({\bf k},\omega)$ at large 
momentum transfer have revealed several features unexpected from the
point of view of weakly correlated fermion systems: 
(a) the very different
form of $N({\bf k},\omega)$ compared to the spin response function
$S({\bf k},\omega)$, which share common features in usual fermionic
systems;
(b) a strong
suppression of low energy $2k_F$ scattering in the density response,
(c) a broad incoherent peak in the higher energy range $t\leq\omega \leq 8t$
with large dispersion,
whose shape is rather insensitive to hole concentration  ($\delta=0.1$ and $0.25$)
and exchange interaction $J$.   
Furthermore, exact diagonalization studies show 
(d) a low-energy structure on scale $J$
for certain momenta which grows in intensity with $J$.

While considerable analytical work has been done to explain the spin
response of the $t$-$J$ model\cite{fuk}, only few authors analyzed 
$N({\bf k},\omega)$. Wang {\it et al.}\cite{wan91}
 studied collective excitations in the
density channel and found sharp peaks at large momenta
corresponding to free bosons. Similar results were obtained by Gehlhoff and
Zeyher\cite{geh95} using the $X$-operator formalism. In these calculations 
fluctuations were treated in the leading order of a $1/N$-expansion 
which is however not sufficient to explain the incoherent character of
 $N({\bf k},\omega)$. 
It was shown subsequently by Khaliullin and Horsch~\cite{KH-H},
 who used a slave-boson approach, 
that the next order in $1/N$-expansion is crucial to obtain renormalized 
polaron-like boson-propagators 
for the holes, which explained both the incoherent structure 
at high energy and the appearance
of a low energy peak associated with the coherent polaron motion. 
The characteristic energy 
scale of the low-energy polaron peak is $\tilde{J}\simeq \chi J+\delta t$, 
where $\chi$ is 
the uniform RVB-order parameter relevant for the overdoped regime studied 
in that work. 
In addition on the same energy scale $\tilde{J}$ there is a smooth 
Fermi-liquid like particle-hole continuum, which reflects the Fermi surface, 
with total weight  $\sim \delta^2$. 
Further studies \cite{V-B,lee96} analysed
the incoherent part of $N({\bf k},\omega)$ yet did not obtain the
low-energy  structure.

The aim of the present work is to investigate  $N({\bf q},\omega)$
in the low-doping regime.  
We confine our study to the AF 
long-range ordered phase which implies a different Fermi surface and 
collective spin wave excitations. There are two key questions: (i) what theory
is needed to produce spectra at high energy similar to the large doping regime
as required by diagonalization studies and (ii) what are the implications
of the long-range magnetic order and the small hole-pocket Fermi surface 
for the low energy charge response?

\section{MODEL AND DENSITY RESPONSE}

We consider here the $2D$ $t-J$ model as  the generic model for the
description of strongly correlated electrons in cuprates:
\begin{equation} H_{tJ}= -t\sum_{\langle i,j\rangle\sigma}^{}
(\tilde c_{i\sigma}^{+}\tilde c_{j\sigma} +\mbox{H.c.})+
J\sum_{\langle i,j\rangle}^{}{\bf S}_{i}{\bf S}_{j} ,
\label{t-J}
\end{equation}
where the operator $\tilde c_{i\sigma}=c_{i\sigma}(1-n_{i\bar\sigma})$
describes the annihilation of an electron with spin $\sigma$ at site $i$
with the  constraint of no double occupancy, ${\bf S}_{i}$ is the spin
operator, and the summation $\langle i,j\rangle $ runs over 
nearest-neighbor  bonds on the square lattice.

We rewrite the constrained electron operators in terms of spinless 
fermions $f_{i}$ which denote the holes and use Holstein-Primakoff 
bosons $b_{i}$ to describe the spin degrees of freedom.
After transformation to momentum representation 
the Hamiltonian $H_{tJ}=H_t+H_J$ takes  the form
\begin{equation}
H_{t}=zt\sum_{{\bf k},{\bf q}}^{} f_{\bf k-q}^{+}f_{\bf k}
(\gamma_{\bf k-q}b_{\bf q}^{+}+\gamma_{\bf k}b_{-\bf q})
+ \mbox{ H.c.},
\label{Ht}
\end{equation}
\begin{equation}
H_{J}=\frac{J}{2}\sum_{\bf q}^{}
[\gamma_{\bf q}(b_{\bf q}^{+}b_{-\bf q}^{+}+b_{-\bf q}b_{\bf q}) +
2b_{\bf q}^{+}b_{\bf q}],
\label{HJ}
\end{equation}
\begin{figure}[htb]
\mbox{}\\
\hspace*{0.5cm}
\vspace*{0cm}
\includegraphics[scale=0.4]{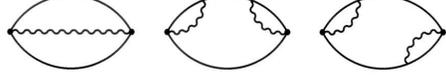}


\vspace*{-0.8cm}
\caption{ Diagrams included in the calculation of the memory function.
Solid and wavy lines represent renormalized spinless fermion 
and  magnon propagators, respectively. 
}
\label{FIG0}
\end{figure}
\begin{figure}
\mbox{}\\
\epsfysize=6.0in
\hspace*{0.5cm}
\includegraphics[scale=0.3]{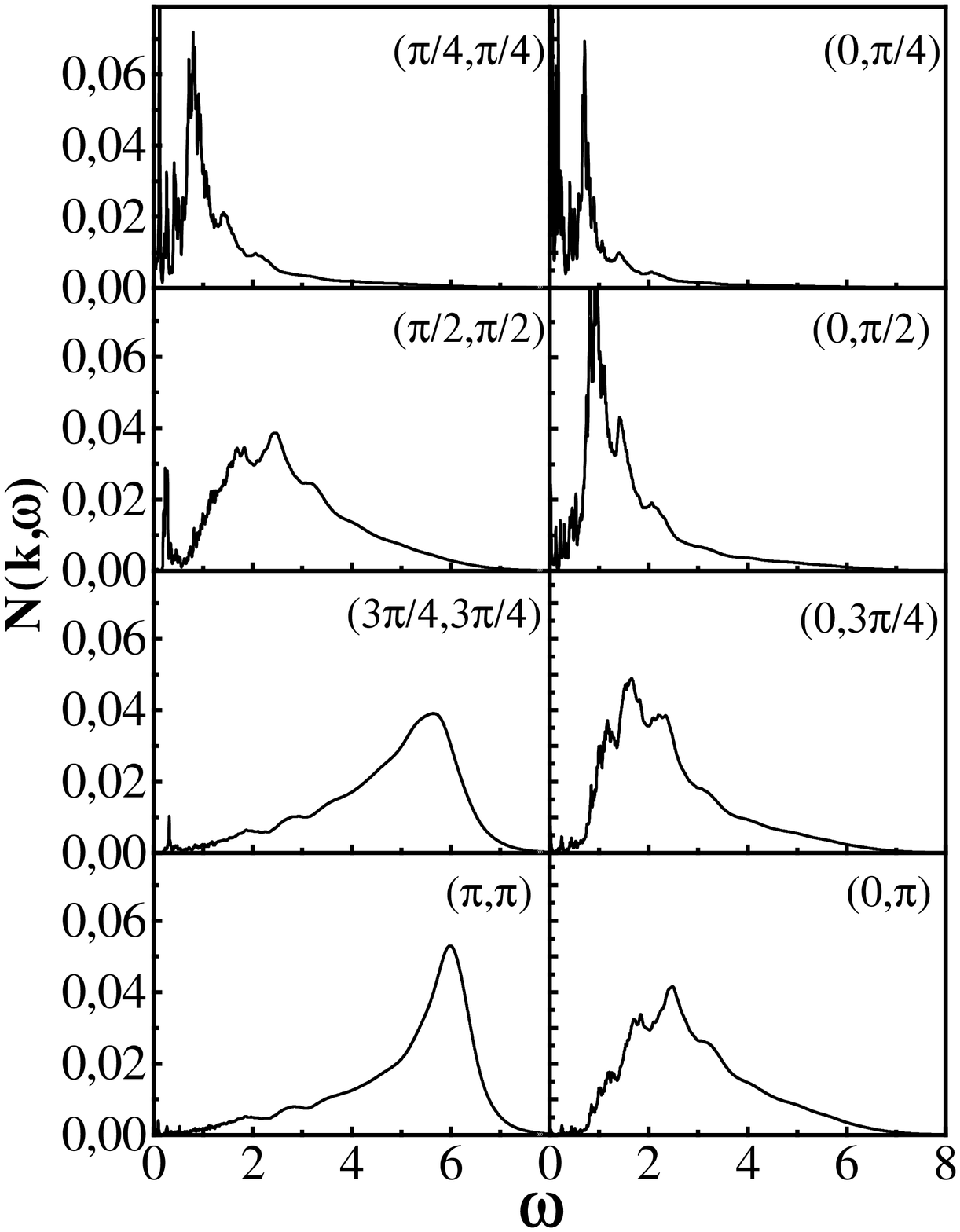}

\vspace*{-0.8cm}
\caption{Density fluctuation spectra for $\delta=0.1$  calculated via 
the slave-fermion approach for $J=0.4$ and momenta along $(\pi,\pi)$ 
and $(\pi,0)$ directions.
The energy unit is $t=1$. }
\label{FIG1}

\end{figure}
where $\gamma_{\bf k}=0.5(\cos k_{x}+\cos k_{y})$ while
$z=4$ denotes the number of nearest neighbours.
This Hamiltonian has the structure of a polaron problem where spinless fermions
(holes)
are coupled to collective spin excitations\cite{M_H}.

The density fluctuation spectrum
\begin{equation}
N({{\bf k},\omega})=-\frac{1}{\pi}\mbox{Im}\chi({\bf k},\omega),
\label{nwq}
\end{equation}
where 
$\chi({\bf k},\omega)=\langle\langle n_{\bf k}\mid n_{-{\bf k}}
\rangle\rangle_{\omega}$ and
$n_{\bf k}=\sum_{i,\sigma}^{}\tilde c_{i,\sigma}^{+}
        \tilde c_{i,\sigma}^{}\mbox{e}^{i{\bf k}{\bf R}_{i}},$
is defined via the density correlation function 
\begin{equation}
\langle\langle n_{\bf k}\mid n_{-{\bf k}}\rangle\rangle_{\omega}=
-i\int\limits_{0}^{\infty}\mbox{e}^{i\omega t}
\langle [n_{\bf k}(t),n_{\bf k}(0)]\rangle dt .
\end{equation}
Calculation of $\chi({\bf k},\omega)$ by expansion in a small
parameter usually fails due to the singular character of this quantity 
for small $\omega$.
This problem can be resolved by rewriting  
$\chi({\bf k},\omega)$ in terms 
of the static susceptibility  $\chi_{0}({\bf k})$ and a selfenergy  
$M({\bf k},\omega)$\cite{MFM,plakida}:
\begin{equation}
\chi({\bf k},\omega)=\chi_{0}({\bf k})\frac{M({\bf k},\omega)}
{\omega+M({\bf k},\omega)} .
\end{equation}
The evaluation of the memory function  $M({\bf k},\omega)$ by 
expanding in a small parameter (here the coupling constant $t$)
gives correct results for  $\chi({\bf k},\omega)$  valid for all frequencies.
From this expression it is obvious that an evaluation of $M$ to low 
order in the coupling constant $t$ implies for the charge susceptibility
$\chi$ a partial resummation up to infinite order.
We shall find that this form is crucial for the proper description of
the charge excitations at high energy, which are of collective nature
and induced by the denominator $\omega + M$ which can become small.
In second order approximation in $t$ we can write:
\begin{equation}
\omega\chi({\bf k},\omega) \simeq \chi_{0}({\bf k})M({\bf k},\omega),
\end{equation}
and determine the memory function by a low order calculation of $\chi$.
Repeated application of equation of motions yields
\begin{equation}
M({{\bf k},\omega})=[f({\bf k},\omega)-f({\bf k},0)]/\chi_{0}({\bf k}),
\label{chikom}
\end{equation}
where $f({\bf k},{\omega})=
-\langle\langle \dot n_{{\bf k}}\mid \dot n_{-{\bf k}}\rangle\rangle_{\omega} $. 
The diagrammatic representation of the  $f({\bf k},{\omega})$ function
is shown in Fig. 1.
The fermion propagators are evaluated within selfconsistent Born
approximation at finite doping concentration;
more details will be given elsewhere.
The still unknown  $\chi_{0}({\bf k})$ is determined
from the integral relation
\begin{equation}
 N({\bf k}) = \int\limits_{}^{} N({\bf k},\omega)d\omega ,
\label{eq29}
\end{equation}
and the fact that the static charge structure factor $N({\bf k})$ is well
approximated in the strong correlation regime by the $N({\bf k})$ for 
spinless fermions\cite{Putikka}.

The density fluctuation spectra obtained by the slave-fermion calculation
are displayed in Fig.2. $N({\bf k},\omega)$ is characterized by a broad
continuum which moves to higher energy as $k$ increases. At $(\pi,\pi)$
the spectrum peaks at $\omega\sim 6 t$, i.e., the variation  
of $N({\bf k},\omega)$ with $k$ and the energy scale are fully consistent with
the diagonalization results for small systems\cite{toh95}.
Low energy structures on scale $J$, due to quasiparticle scattering between 
hole pockets, appear at small and intermediate momentum transfers and 
vanish at the zone boundary. The pronounced low-energy structure at large
momentum transfer in $(\pi,0)$ direction found in the spin liquid regime
\cite{KH-H} is absent. 

\section{BOND-STRETCHING PHONONS}

The high energy planar breathing and bond-stretching oxygen modes 
in $(\pi,\pi)$ and $(\pi,0)$ direction, respectively, show
the strongest doping dependence from all phonon modes ~\cite{Pintsch}
which appears to be a generic feature of cuprates.
Inelastic neutron scattering studies of the LO mode in 
$(\pi,0)$ direction in La$_{2-x}$Sr$_x$CuO$_4$ ($x=0.15$)\cite{Pintsch,neutron}
reveal a $\sim 10 \%$ renormalization near $(\pi,0)$ as compared to the undoped
system, while there is no significant renormalization 
at the $\Gamma$-point.
Surprisingly the linewidth assumes a maximum at intermediate 
$\bf k$\cite{neutron}.

The coupling of these phonon modes to density fluctuations is described by
\cite{Pekin} 
\begin{equation}
H_{e-ph}=g\sum_{i}^{}f_{i}^{+}f_{i}
(u_{x}^{i}-u_{-x}^{i}+u_{y}^{i}-u_{-y}^{i}),
\label{e-ph}
\end{equation}
where $u_{x}^{i}, u_{y}^{i}$ are the displacements of the four oxygen
neighbors of the hole, $g$ is the electron-phonon coupling constant.
This coupling originates from the change of the
Zhang-Rice singlet energy by breathing and bond-stretching modes.

The phonon Green's function has the following  form for ${\bf k}=(k,0)$
\begin{equation}
D(k,\omega)=\frac{\omega_{0}}{\omega^{2}-\omega_{0}^{2} (1+4\xi
\mbox{sin}^{2}
\frac{k}{2} \chi (k,\omega))}  ;
\end{equation}
here we assume a k-independent bare phonon frequency
$\omega_0=0.3$ and a normalized coupling constant  
$\xi=g^2/Kt=0.25 $\cite{Pekin}.
For momenta ${\bf k}=(k,k)$ the constant $4\xi$  in Eq.(11) 
should be replaced by $8\xi $. The phonon spectral function 
$B(k,\omega)=-\frac{1}{\pi}Im D(k,\omega)$ is displayed in Fig. 3.
\begin{figure}[htb]
\noindent
\hspace*{1cm}
\epsfysize=6.0in
\includegraphics[scale=0.43]{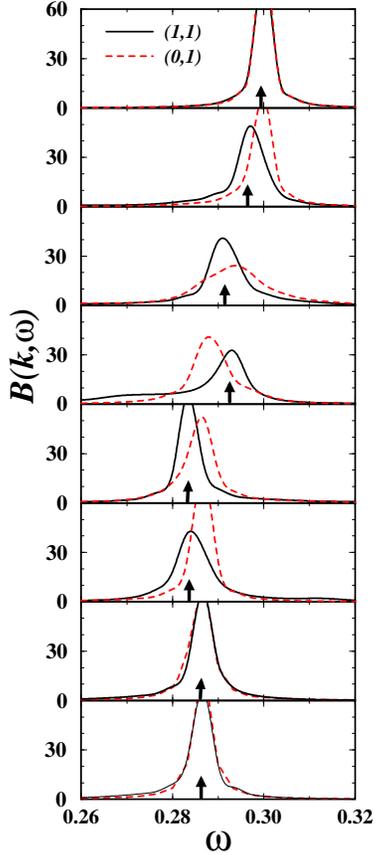}
\vspace*{-0.6cm}
\caption{Phonon spectral function 
at $\delta=0.1$ for different momenta along
$(k,0)$ and $(k,k)$ directions, where $k=\pi/8$ for the top panel
and increases in steps of $\pi/8$ in downward direction.
Here $\omega_0=0.3$; arrows indicate $\omega_k$ along $(k,k)$. }
\label{FIG3}
\end{figure}
Figure 3 shows a large increase of the renormalization 
of $\omega_{\bf k}$ with increasing $k$,
which is similar for the two directions, 
with rather abrupt changes in  $\omega_{\bf k}$ at intermediate $k$.
The damping appears to be
largest not at the zone-boundary but at intermediate $k$. 
We note that earlier
calculations for larger doping, $\delta=0.15-0.25$, show a significantly 
larger energy change at $(\pi,0)$ than at $(\pi,\pi)$, which was due to the 
pronounced low energy polaron related structure in $N({\bf k},\omega)$
near $(\pi,0)$ 
in the spin disordered state\cite{Pekin}. 
The low-energy structure of   $N({\bf k},\omega)$ in 
the AF ordered system considered here is markedly
different, and suggests nontrivial changes for the phonon renormalization
as function of doping. 

Hence we conclude by stressing that phonon modes, and in particular those
discussed here, provide a sensitive probe for the anomalous low-energy
charge response of doped Mott insulators.
Furthermore it is expected that measurements at different doping concentrations
will reveal significant changes of phonon energies and damping due to the 
changes in the spin response and in the topology of the Fermi surface, 
which in turn strongly affect the charge response at
low energy.


\begin{thebibliography}{9}

\bibitem{toh95} T. Tohyama, P. Horsch, and S. Maekawa,
  Phys. Rev. Lett.  74 (1995) 980.

\bibitem{fuk} H. Fukuyama, H. Kohno, B. Normand, and T. Tanamoto,
 Physica B 213\&214 (1995) 6  and references therein.

\bibitem{wan91} Z. Wang, Y. Bang, and G. Kotliar, Phys. Rev. Lett. 
 67 (1991) 2733.

\bibitem{geh95} L. Gehlhoff and R. Zeyher, Phys. Rev. 52 (1995) 4635.

\bibitem{KH-H}  G.~Khaliullin and P.~Horsch,
Phys.~Rev.~B  54 (1996) R9600 .

\bibitem{V-B} M.~Vojta and K.~Becker, Eur.Phys.J. B 3 (1998) 427.

\bibitem{lee96}
D.~K.~K.~Lee, D.~H.~Kim, and P.~A.~Lee,
Phys.~Rev.~Lett. 76 (1996) 4801.

\bibitem{M_H} G.~Mart\'inez and P.~Horsch,
Phys.~Rev.~B.~44 (1991) 317 and references therein.

\bibitem{MFM} W.~G\"otze and P.~W\"olfle,
Phys.~Rev.~B. 6 (1972) 1226.

\bibitem{plakida} G. Jackeli and N.M. Plakida, Phys.Rev. 60 (1999) 5266.

\bibitem{Putikka} W.O. Putikka, R.L. Glenister, R.R.P. Singh,
and H. Tsunetsugu,
Phys. Rev. Lett. 73 (1994) 170.

\bibitem{Pintsch} L.~Pintschovius, and W.~Reichardt,
Physical Properties of HTSC IV, edited by D.~Ginsberg (World
Scientific, Singapore, 1994), p.344 and 347.

\bibitem{neutron} McQueeney {\it et al.}, Phys. Rev. Lett. 82 (1999) 628; 
L. Pintschovius and M. Braden, Phys. Rev. B 60 (1999) R15039. 

\bibitem{Pekin}  G.~Khaliullin and P.~Horsch,
Physica C 282 (1997) 1751.

\end{thebibliography}
\end{document}